\documentclass[aps,a4paper,superscriptaddress,nofootinbib,twocolumn,10pt]{revtex4-1}
\usepackage[paperwidth=21cm,paperheight=29.7cm,top=2.54cm,bottom=2.54cm,left=2cm,right=2cm]{geometry}
\usepackage{amsmath,amssymb,bm,latexsym,acronym,float,soul,nicefrac,verbatim,cancel}
\usepackage{multirow,dcolumn,longtable,footnote,appendix,graphicx,color,subfigure}
\usepackage[colorlinks,linkcolor=blue,citecolor=blue,urlcolor=blue ]{hyperref}
\usepackage[normalem]{ulem}
\newcommand{\SPA}{School of Physics and Astronomy, Sun Yat-sen University (Zhuhai Campus), Zhuhai 519082, China}
\allowdisplaybreaks
\newcommand{\be}{\begin{equation}}
\newcommand{\ee}{\end{equation}}
\newcommand{\bea}{\begin{eqnarray}}
\newcommand{\eea}{\end{eqnarray}}
\newcommand{\nn}{\nonumber}
\newcommand{\pd}{\partial}
\newcommand{\td}{\tilde}

\newcommand{\tdg}{\td{g}}

\newcommand{\cA}{{\cal A}}
\newcommand{\cB}{{\cal B}}

\newcommand{\cE}{{\cal E}}
\newcommand{\cF}{{\cal F}}

\newcommand{\cO}{{\cal O}}
\newcommand{\cP}{{\cal P}}

\newcommand{\cT}{{\cal T}}
\newcommand{\cU}{{\cal U}}
\newcommand{\cV}{{\cal V}}
\newcommand{\cW}{{\cal W}}
\newcommand{\cX}{{\cal X}}

\begin{document}

\title{A hydrodynamical description of gravitational waves}

\author{Jianwei Mei}
\email{Email: meijw@sysu.edu.cn}
\affiliation{\SPA}

%\input{git_tag.tex}
%\date{\commitDATE; \commitIDshort-\commitSTATUS}
%\date{\today}

\begin{abstract}
It is easy to reason that gravity might be the effect of a fluid in disguise, as it will naturally arise in emergent gravity models where gravity is due to the effect of some fundamental particles, with the latter expected to behave collectively like a fluid at the macroscopic scale. We call this the fluid/gravity equivalence. The key difficulty with the fluid/gravity equivalence is to find the correct metric-fluid relation (the relation between the emergent metric and the fluid properties) so that the fluid not only has physically acceptable properties but also obeys the usual hydrodynamic equations, while at the same time the emergent metric also obeys the Einstein equations. Faced with the problem, we have previously made a tentative proposal of the metric-fluid relation, focusing only on obtaining physically acceptable predictions on the fluid properties. In this paper, however, we find that for the general gravitational wave spacetime near the null infinity, the underlying fluid not only has physically acceptable properties, but also satisfies the expected relativistic hydrodynamic equations in the Minkowski background, thus providing a concrete example satisfying both of the major requirements expected for the fluid/gravity equivalence.
\end{abstract}



%%%%%%%%%%%%%%%%%%%%%%%%%%%%%%%%%%%%%%%%%%%%%%%%%%%%%%%%%%%%%%%%
\maketitle
%%%%%%%%%%%%%%%%%%%%%%%%%%%%%%%%%%%%%%%%%%%%%%%%%%%%%%%%%%%%%%%%
\acrodef{GR}{general relativity}
\acrodef{GW}{gravitational wave}
\acrodef{FFC}{{\it Fundamental Frame Conjecture}}
\acrodef{FRW}{Friedmann-Robertson-Walker}
%%%%%%%%%%%%%%%%%%%%%%%%%%%%%%%%%%%%%%%%%%%%%%%%%%%%%%%%%%%%%%%%

\section{Introduction}\label{sec:intro}

The possible connection between gravity and hydrodynamics has been noticed for a long time \cite{Hawking:1972hy,Hartle:1973zz,Hartle:1974gy}. Early works linking black holes to acoustic ``dumb holes" have led to the development of analogue gravity \cite{Unruh:1980cg,Visser:1993ub,Visser:1997ux,Visser:1998qn,Barcelo:2005fc}, which uses the similarities between gravity and matter systems to test ideas about gravity. However, it has proved difficult to reinvent the full complexity of \ac{GR} along this route. Gauge/gravity duality is by far the most powerful in establishing the connection between gravity and hydrodynamics, by providing it with profound conceptual foundations \cite{Damour:1979wya,1982mgm..conf..587D,Price:1986yy,Thorne:1986iy,Policastro:2001yc,Son:2007vk,Gourgoulhon:2005ch,Gourgoulhon:2005ng, Bhattacharyya:2007vjd,Bhattacharyya:2008mz,Bhattacharyya:2008kq,Eling:2008af,Eling:2009pb,Eling:2009pb,Eling:2009sj,Fouxon:2008tb,Emparan:2009cs,Emparan:2011hg,Padmanabhan:2010rp,Hubeny:2011hd}. The essence of the duality treatment is to use boundary hypersurfaces, such as black hole horizons, AdS boundaries, Rindler horizons, and cutoff manifolds \cite{Gourgoulhon:2005ch}, to probe the bulk spacetime, then Navier-Stokes type equations can be obtained in the hydrodynamic limit as boundary data constraints governing the dynamics of some putative dual fluid living on the boundary hypersurfaces \cite{Damour:1979wya,1982mgm..conf..587D,Price:1986yy,Bredberg:2010ky,Bredberg:2011xw,Bredberg:2011jq, Compere:2011dx,Compere:2012mt,Eling:2008af,Eling:2009pb,Eling:2009pb,Eling:2009sj,Eling:2012ni,Gourgoulhon:2005ch,Gourgoulhon:2005ng,Bhattacharyya:2008kq,Kovtun:2004de,Padmanabhan:2010rp,Nakayama:2011bu,Cai:2011xv,Huang:2011kj, Brattan:2011my,Anninos:2011zn,Marolf:2012dr,Pinzani-Fokeeva:2014cka,Lysov:2017cmc,De:2018zxo,Bhattacharyya:2019mbz,Ferreira-Martins:2021cga}. In such treatment, the boundary hypersurfaces are manifestly part of the spacetime, but the dual fluid is more a mathematical tool than a real physical entity that exists in the real world (see, e.g., \cite{Pinzani-Fokeeva:2014cka} for some discussion).

This paper is concerned with a possible connection between gravity and hydrodynamics in the more direct sense. The construction is rooted in ideas from emergent gravity (see, e.g., \cite{Hu:2009jd,Sindoni:2011ej,Carlip:2012wa,Linnemann:2017hdo}), which is concerned with the possibility that part or all aspects of gravity are emergent from some more fundamental structures. For a large class of models, such fundamental structure is nothing but a system of fundamental particles (see, e.g., \cite{Hu:2009jd,Sindoni:2011ej}). Such fundamental particles are expected to behave collectively like a fluid at the macroscopic scale. So one can expect a kind of {\it fluid/gravity equivalence}, in the sense that gravity is equivalent to and thus can be replaced by the effect of some real physical fluid living in the real physical world, in the limit when the fluid approximation is valid. I shall call such fluid hidden fluid because it shows itself only in the disguise of gravity and in a field theory construction it would belong to a hidden sector. Note hidden fluid lives in the same spacetime dimension as gravity and so is fundamentally different from the concept of dual fluid in the above mentioned duality treatment.

Although being conceptually very simple, the fluid/gravity equivalence is facing a fundamental difficulty: without a complete construction of the correct emergent gravity theory at the fundamental level, the metric-fluid relation (the relation between the emergent metric and the properties of hidden fluid) is {\it a priori} not known. Such a relation should be consistent with two major requirements expected for the fluid/gravity equivalence:
\begin{description}
\item[R1] For all physically acceptable solutions to the Einstein equations,{
%%%
\bea \td{R}_{\mu\nu}-\frac12\td{R}\tdg_{\mu\nu}=8\pi G_N\td{T}_{\mu\nu}\,,\label{eq.Ruv}\eea
%%%
where $\td{R}_{\mu\nu}$ and $\td{R}$ are the Ricci tensor and Ricci scalar of the emergent metric $\td{g}_{\mu\nu}\,$, respectively, $\td{T}_{\mu\nu}$ is the stress-energy tensor of matter (plus the contribution of a cosmological constant if needed),} the properties of hidden fluid predicted by the metric-fluid relation should be physically acceptable. This is a highly non-trivial requirement considering the diversity of physically possible spacetimes.
\item[R2] The leading order dynamics of hidden fluid should be governed by the usual hydrodynamic equations in the Minkowski background,
%%%
\bea\pd_\mu\cT^{\mu\nu}=0\,,\label{eq.Tuv1}\eea
%%%
where $\cT^{\mu\nu}$ is the stress-energy tensor of hidden fluid. This is also highly non-trivial because the metric-fluid relation will tie the properties of hidden fluid to the emergent metric, while the latter has to satisfy (\ref{eq.Ruv}).
\end{description}

To search for the correct metric-fluid relation, the best one can do now is to make conjectures on the relation and then to check it against the above requirements.
This is obviously a very difficult task: it is already difficult enough just to find a metric-fluid relation that can satisfy either {\bf R1} or {\bf R2} alone, let along finding one that can satisfy both simultaneously.
With such understanding in mind, we have made a tentative proposal on the metric-fluid relation and have focused on checking the relation against the requirement {\bf R1} only \cite{Mei:2022ksw}.

In this paper, however, when using the proposed relation to study the properties of hidden fluid underlying the general \ac{GW} spacetimes near the null infinity, we find that not only {\bf R1} but also {\bf R2} are satisfied.
This is very surprising because one would naturally expect {\bf R2} to be too far out of reach, especially when {\bf R1} is still being checked. But as we see later in the discussion, the mathematics involved is rather simple and straightforward.

Although such intriguing result does not necessarily mean that we have found the correct metric-fluid relation, it does provide a concrete example satisfying both of the major requirements expected for the fluid/gravity equivalence. This could serve as a good starting point for further studies on the problem. So we would like to report the result here.

The rest of the paper is organized as following. The basic ideas and formulae of the hidden fluid model are recalled in section \ref{sec:HF}. The general \ac{GW} solution in the hidden fluid model is obtained in section \ref{sec:gw}. The hidden fluid underlying the \ac{GW} spacetime is shown to satisfy both (\ref{eq.Ruv}) and (\ref{eq.Tuv1}) in section \ref{sec:hydro}. A conclusion is in section \ref{sec:con}.

\section{The hidden fluid model}\label{sec:HF}

The metric-fluid relation proposed in \cite{Mei:2022ksw} has been motivated from two different considerations.
Firstly, {since the metric-fluid relation is {\it a priori} not known, one may try to start with the most general possibility, written schematically as}
%%%
\bea\cT^{\mu\nu}=\cF^{\mu\nu}[\tdg,\pd\tdg,\pd\pd\tdg\,,\cdots]\,,\label{def.tdguv}\eea
%%%
where $\cF^{\mu\nu}[\cdots]$ is a {completely general} functional of the emergent metric and its derivatives. {To determine the structure of $\cF^{\mu\nu}[\cdots]\,$, we note the following: (i) as a preliminary attempt, it is reasonable to start with cases in which $\cF^{\mu\nu}[\cdots]$ is an analytic function, so by definition it can be expanded in terms of the polynomials of its arguments; (ii) it is natural to start with neglecting the derivative terms, since derivative expansion, long wavelength limit and hydrodynamic limit often come hand in hand; (iii) it is also reasonable to require that $\cF^{\mu\nu}[\cdots]$ transforms covariantly under general coordinate transformations. These choices and requirements seem to lead to a unique solution to (\ref{def.tdguv}),}
%%%
\bea\cT^{\mu\nu}=-\varrho_0~\tdg^{\mu\nu}\,,\label{def.hf1}\eea
%%%
{where $\varrho_0$ is some scalar function.}
Secondly, a relation of the form (\ref{def.hf1}) also arises in concrete physical models, such as in certain emergent gravity models \cite{Kiritsis:2014yqa,Betzios:2020sro} and the contribution of the cosmological term to Einstein equations. As such, (\ref{def.hf1}) has been conjectured as {\it the} metric-fluid relation, with $\varrho_0$ taken to be a constant and identified as the average energy density of hidden fluid at the present time of the universe.

By talking about hidden fluid in the context of emergent gravity, one typically assumes that there is a flat background spacetime on which the fundamental theory of hidden fluid can be defined. Such requirement singles out a particular set of coordinate systems, called the fundamental frames, which are unique up to a Poincar\'{e} transformation. (To compare with other coordinate choices, I shall also call such coordinate choice the fundamental gauge.) Since in \ac{GR} one can work in arbitrary coordinate systems, one must find a way to identify the fundamental frames. Based on a universality consideration, it has been conjectured \cite{Mei:2022ksw} that the fundamental frames should be ones in which the emergent metric can be written in the extended Kerr-Schild form \cite{Llosa:2004uf,Llosa:2008yk,Harte:2014ooa},\footnote{Throughout this work, indices on tensors with a tilde are raised and lowered using the emergent metric, while those on tensors without a tilde are raised and lowered using the Minkowski metric.}
%%%
\bea \tdg_{\mu\nu}&=&\frac1\cA\eta_{\mu\nu}+\Big(\frac1\cA-\frac1\cB\Big) (\cU_\mu\cU_\nu-\cV_\mu\cV_\nu)\,,\nn\\
\tdg^{\mu\nu}&=&\cA\eta^{\mu\nu}+(\cA-\cB)(\cU^\mu\cU^\nu-\cV^\mu\cV^\nu)\,,\label{def.hf2}\eea
%%%
where $\eta_{\mu\nu}$ is a Minkowski metric, $\cA$ and $\cB$ are scalars, and $\cU$ and $\cV$ are vectors,
%%%
\bea \cU\cdot\cU=-\cV\cdot\cV=-1\,,\quad \cU\cdot\cV=0\,.\label{uv.constraints}\eea
%%%
{We will use the most plus convention for all the metrics in this paper, so $\cU$ is timelike and $\cV$ is spacelike.}

Combining (\ref{def.hf1}) and (\ref{def.hf2}), one has for the stress-energy tensor of hidden fluid,
%%%
\bea\cT^{\mu\nu}=\cE\cU^\mu\cU^\nu + \cP \Pi^{\mu\nu}+\cX^{\mu\nu}\,,\label{def.Tab}\eea
%%%
where $\cU^\mu$ is the four velocity, $\cE=\varrho_0\cB$ is the energy density, $\cP=-\frac13\varrho_0(2\cA+\cB)$ is the pressure,
%%%
\bea\cX^{\mu\nu}=\varrho_0(\cA-\cB)(\cV^\mu\cV^\nu-\frac13\Pi^{\mu\nu})\,,\label{def.cX}\eea
%%%
represents the dissipative part, and $\Pi^{\mu\nu}=\eta^{\mu\nu}+\cU^\mu\cU^\nu$ is the transverse projector. Note the relations, $\cA=-\frac1{2\varrho_0}(\cE+3\cP)\,$, $\cB=\frac{1}{\varrho_0}\cE\,$, and $\eta_{\mu\nu}\cX^{\mu\nu}=\cU_\mu\cX^{\mu\nu}=0\,$.

To fully specify the relation between the properties of hidden fluid and the emergent metric, one not only needs the relation (\ref{def.hf1}), but also needs the fundamental gauge condition (\ref{def.hf2}) and the physical interpretation of the quantities in the stress energy tensor (\ref{def.Tab}). So these relations as a whole constitutes the full metric-fluid relation that we want to use, and we have given it a special name, the {\it hidden fluid model} \cite{Mei:2022ksw}.

Due to the speculative nature of both (\ref{def.hf1}) and the fundamental gauge condition (\ref{def.hf2}), there is huge chance for the model to fail, i.e., with (\ref{def.hf1}) and (\ref{def.Tab}) making unphysical predictions on the properties of hidden fluid. We have tested the model against a few highly symmetric spacetimes, and the properties of hidden fluid underlying these spacetimes have been found to be acceptable \cite{Mei:2022ksw}. Some intriguing results have also been found in the process. For example, the flat \ac{FRW} metric in the fundamental gauge can be written as
%%%
\bea&&ds^2=a(t)^2\Big(-dt^2+dr^2+r^2d\Omega_2\Big)\,,\nn\\
&&a(t)=\sqrt{\varrho_0/\cE(t)}\,,\quad \cE(t)=\cE_i\Big(1-\frac{t}{t_f}\Big)^2\,, \label{metric.FLRW1}\eea
%%%
where $t\in[0\,,\,t_f)\,$, $t_f\equiv\ell/a(0)\,$, $\ell= \sqrt{3/\Lambda}\,$, $\cE_i=\cE(0)\,$, with $\cE(t)$ being the energy density of hidden fluid at time $t\,$ and $\Lambda$ being the cosmological constant. The solution appears to show that, the universe started at $t=0$ with the sudden appearance of hidden fluid, having an initial energy density $\cE_i\,$, then the energy density of hidden fluid started to decrease continuously, leading to a continuous expansion of the universe. The dramatic variation in the energy density of hidden fluid is most likely due to a first order phase transition involving the fabric of space that underlies the Minkowski background. So in the hidden fluid model, there is the intriguing possibility that the origin and the expansion of the universe is intimately related to some phase transitions.

\section{The general gravitational wave solution in the hidden fluid model}\label{sec:gw}

We have turned to \ac{GW} spacetimes in an effort to further test the model against the requirement {\bf R1}, as discussed in the first section. The general \ac{GW} solution in an asymptotically flat background has been known for a long time and is described by the Bondi-Sachs metric \cite{Bondi:1962px,Sachs:1962wk,Sun:2022pvh}. So to study the properties of hidden fluid underlying \ac{GW} spacetimes, one may transform the Bondi-Sachs metric into the form of (\ref{def.hf2}) through a coordinate transformation. We have checked that this can indeed be done. In the following, however, we choose to solve for the \ac{GW} solution in the fundamental gauge directly in order to lay a better foundation for our later discussion.

Working in the fundamental gauge with the coordinates $x^\mu\in\{u,r,\theta,\phi\}\,$, where $u$ is the retarded time, we choose the flat background metric as
%%%
\bea ds_0^2&=&\eta_{\mu\nu}dx^\mu dx^\nu\nn\\
&=&-du^2-2du dr+(r_0+r)^2q_{AB}d\theta^Ad\theta^B\,,\eea
%%%
where $r_0$ stands for the large distance between the wave zone and the \ac{GW} sources, $\theta^A\in\{\theta,\phi\}\,$, and $q_{AB}$ is the fixed round metric on the unit two-sphere,
%%%
\bea q_{AB}d\theta^Ad\theta^B=d\theta^2+\sin^2\theta d\phi^2\,.\eea
%%%
The constraints in (\ref{uv.constraints}) can be solved with
%%%
\bea\cU_1&=&U_1\,,\quad \cU_2=U_2\,,\quad \cV_1=V_1\,,\quad \cV_2=V_2\,,\nn\\
\cU_3&=&(r_0+r)U_3\,,\quad \cU_4=(r_0+r)\sin\theta~U_4\,,\nn\\
\cV_3&=&(r_0+r)V_3\,,\quad \cV_4=(r_0+r)\sin\theta~V_4\,,\label{GW.UV1}\eea
%%%
and
%%%
\bea U_1&=&\frac1{2U_2}(U_2^2+U_3^2+U_4^2+1)\,,\nn\\
V_1&=&\frac1{2V_2}(V_2^2+V_3^2+V_4^2-1)\,,\nn\\
U_2&=&\frac12(F_1-F_2)\,,\nn\\
V_2&=&\frac12(F_1+F_2)\,,\nn\\
V_3&=&\frac1{U_2}(V_2U_3-F_3)\,,\nn\\
V_4&=&\frac1{U_2}(V_2U_4-F_4)\,,\nn\\
F_1&=&-\frac{1}{F_2}(F_3^2+F_4^2)\,.\label{GW.UV2}\eea
%%%
As expected, there are five free functions, ($U_3\,$, $U_4\,$, $F_2\,$, $F_3\,$, $F_4\,$), left in $\cU$ and $\cV\,$.

The general \ac{GW} solution can be found in the limit $r_0\rightarrow\infty\,$, with respect to which the functions are expanded as
%%%
\bea U_\mu&=&\sum_{n=0}^\infty{U_{\mu n}\xi^n}\,,\quad \mu=3,4\,,\nn\\
F_\mu&=&\sum_{n=0}^\infty{F_{\mu n}\xi^n}\,,\quad \mu=2,3,4\,,\nn\\
\cA&=&1+\sum_{n=1}^\infty{A_n\xi^n}\,,\quad
\cB=\cA+\sum_{n=1}^\infty{B_n\xi^n}\,,\label{ansatz}\eea
%%%
where {$\xi\equiv a_0/r_0<<1\,$ is the expansion parameter, while all the coefficient functions are of the order $\cO(1)\,$. We assume that $a_0\,$,} $U_{\mu 0}\,$, $F_{\mu 0}\,$, $A_1$ and $B_1$ depend only on $\{u,\theta,\phi\}\,$, while all higher order coefficient functions depend on all coordinates, $x^\mu\in\{u,r,\theta,\phi\}\,$. For later convenience, I shall refer to $U_{\mu n}$, $V_{\mu n}$, $A_{n+1}$ and $B_{n+1}$ as the order-($n+1$) coefficients, because they contribute to the vacuum Einstein equation (\ref{eq.Ruv}) starting from the order {$\cO(\xi^{n+1})\,$. What's more, whenever we say that a tensor equation $E_{\mu\nu\cdots \sigma}=0\,$ is satisfied to the order $\cO(\xi^n)\,$, we always mean $E_{\mu\nu\cdots \sigma}dx^\mu dx^\nu\cdots dx^\sigma \sim o(\xi^n)\,$, assuming that $du,dr\sim\cO(1)\,$ and $d\theta,d\phi\sim\cO(\xi)\,$.}

The order-1 coefficients, which solve (\ref{eq.Ruv}) to the order {{$\cO(\xi)\,$}}, can be explicitly found as
%%%
\bea U_{30}&=&-(2U_{20}+V_{20}{f_\theta})\,,\nn\\
U_{40}&=&-(2U_{20}+V_{20}{f_\phi})\,,\nn\\
F_{30}&=&\frac{{f_\theta}}{{f_+}}\,,\quad F_{40}=\frac{{f_\phi}}{{f_+}}\,,\nn\\
A_1&=&{f_+}\,,\quad B_1=-2A_1\,,\label{GW.s1}\eea
%%%
where $U_{20}$ and $V_{20}$ are the leading order terms in the expansion of $U_2$ and $V_2\,$, respectively, ${f_\theta}$ and ${f_\phi}$, together with $F_{20}$, are arbitrary functions of $\{u,\theta,\phi\}\,$, and we have defined {$f_+=f_\theta^2+f_\phi^2\,$. We also note $a_0$ can only depend on $(\theta,\phi)\,$.}

The emergent metric at order-1 is
%%%
\bea ds^2=(1-{\xi f_+})ds_0^2-{2\xi f_+}d\ell dk+{\cO(\xi^2)}\,, \label{metric.s0}\eea
%%%
where
%%%
\bea d\ell&\equiv&(d\cU+d\cV)_{\text{leading-order}}\nn\\
&=&\frac{1}{F_{20}{f_+}}\Big\{({f_\theta}+2)d\theta+({f_\phi}+2)\sin\theta d\phi\nn\\
&&-\frac12\Big[({f_\theta}+2)^2 +({f_\phi}+2)^2+1\Big]du-dr\Big\}\,,\nn\\
dk&\equiv&(d\cU-d\cV)_{\text{leading-order}}\nn\\
&=&-F_{20}\Big\{({f_\theta}-2)d\theta+({f_\phi}-2)\sin\theta d\phi\nn\\
&&+\frac12\Big[({f_\theta}-2)^2 +({f_\phi}-2)^2+1\Big]du+dr\Big\}\,.\nn\\
\eea
%%%
It is remarkable that although (\ref{GW.s1}) leaves three functions ($F_{20}\,$, ${f_\theta}$ and ${f_\phi}\,$) undetermined, only ${f_\theta}$ and ${f_\phi}$ are physically relevant, while $F_{20}$ is a scaling freedom in $d\ell$ and $dk\,$. This is consistent with the fact that \acp{GW} in \ac{GR} only have two polarizations.

{From (\ref{metric.s0}), one can see that the magnitudes of \acp{GW} are of the order $\cO(\xi)$, while the components of $\cU$ are only determined by $F_{20}\,$, ${f_\theta}$ and ${f_\phi}\,$, and are of the order $\cO(1)\,$, indicating that} the velocity of hidden fluid is not controlled by the magnitude of \acp{GW}: hidden fluid can have order $\cO(1)$ velocities no matter how weak the \acp{GW} are. Considering the countless number of \ac{GW} sources in the universe, the motion of hidden fluid can be extremely chaotic. This behavior might be surprising but since the velocity of hidden fluid is always timelike, see (\ref{uv.constraints}), and due to (\ref{def.hf1}), the stress-energy tensor of hidden fluid is always conserved in the emergent spacetime,
%%%
\bea\td\nabla_\mu\cT^{\mu\nu}=0\,,\label{eq.Tuv2}\eea
%%%
where $\td\nabla_\mu$ is defined with the emergent metric, so no physical principles have been violated.

Contrary to the velocities, the {variation in the} energy density and the pressure of hidden fluid do share the perturbative nature of weak \acp{GW},
%%%
\bea \cE&=&+\varrho_0\Big[1-{\xi f_+}+{\cO(\xi^2)}\Big]\,,\nn\\
\cP&=&-\varrho_0\Big[1+{\frac{\xi f_+}3}+{\cO(\xi^2)}\Big]\,.\label{rst.cEcP}\eea
%%%
Together with (\ref{def.hf2}) and (\ref{def.cX}), one can see that it is the cancellation between the energy density and pressure that makes the effect of hidden fluid weak, despite the fact that it has large velocities.

To finish this section, we note the order-2 equations do not place any extra constraint on the order-1 coefficients that have been found so far.

\section{The hydrodynamic equations for gravitational waves}\label{sec:hydro}

By construction, the conservation of $\cT^{\mu\nu}$ in the emergent spacetime is automatically satisfied as shown in (\ref{eq.Tuv2}). So the dynamics of hidden fluid is totally governed by the Einstein equations (\ref{eq.Ruv}), through the relation (\ref{def.hf1}). We would like to know what this means for the hidden fluid underlying the \ac{GW} spacetimes.

In the usual study of \acp{GW}, one perturbs the emergent metric as
%%%
\bea\tdg_{\mu\nu}\approx\eta_{\mu\nu}+h_{\mu\nu}\,,\label{ansatz.GW}\eea
%%%
and linearize the vacuum Einstein equation (\ref{eq.Ruv}) as,
%%%
\bea \Box h_{\mu\nu}-\pd_\rho(\pd_\mu\check{h}_\nu^\rho+\pd_\nu\check{h}_\mu^\rho)=0\,,\label{eq.huv}\eea
%%%
where $\Box=\eta^{\rho\sigma}\pd_\rho\pd_\sigma\,$, $\check{h}_{\mu\nu}=h_{\mu\nu}-\frac12\eta_{\mu\nu}h$ and $h=\eta^{\mu\nu}h_{\mu\nu}\,$. By imposing the harmonic gauge condition,
%%%
\bea\pd_\mu\check{h}^\mu_\nu=0\,,\label{eq.huv1}\eea
%%%
one can obtain from (\ref{eq.huv}) the familiar wave equations for \acp{GW},
%%%
\bea\Box h_{\mu\nu}=0\,.\label{eq.huv2}\eea
%%%

In the hidden fluid model, however, the general \ac{GW} solution has to be found in the fundamental gauge. Here a remarkable feature arises: simply by using the fundamental gauge and the general ansatz (\ref{ansatz}), the would-be wave equation (\ref{eq.huv2}) is automatically satisfied at the order {$\cO(\xi)\,$}, i.e.,
%%%
\bea (\Box h_{\mu\nu})dx^\mu dx^\nu&\sim&{\cO(\xi^2)}\,.\label{boxhuv}\eea
%%%
As a result, (\ref{eq.huv}) reduces at the order {$\cO(\xi)\,$} to
%%%
\bea \pd_\rho(\pd_\mu\check{h}_\nu^\rho+\pd_\nu\check{h}_\mu^\rho)=0\,,\label{eq.huv3}\eea
%%%
which is then solved by the order-1 solution (\ref{metric.s0}). In fact, (\ref{metric.s0}) not only solves (\ref{eq.huv3}) but also solves (\ref{eq.huv1}) to the order {$\cO(\xi)\,$}. So the fundamental gauge is in some sense opposite to the harmonic gauge: In the harmonic gauge, (\ref{eq.huv1}) is the gauge condition and (\ref{eq.huv2}) is the dynamical equation, while in the fundamental gauge, (\ref{eq.huv2}) is satisfied through a gauge choice and (\ref{eq.huv1}) is solved as the dynamical equation.

Now for the general \ac{GW} solution (\ref{metric.s0}), one can find that $h=0\,$ at the order {$\cO(\xi)\,$}, and so from (\ref{def.hf1}),
%%%
\bea\check{h}^{\mu\nu}\approx\eta^{\mu\nu}+\frac1{\varrho_0}\cT^{\mu\nu}+{\cO(\xi^2)}\,.\eea
%%%
Plugging into (\ref{eq.huv1}), one immediately gets (\ref{eq.Tuv1}) at the ${\cO(\xi)}$ order. So for hidden fluid underlying the \ac{GW} spacetime, (\ref{eq.Tuv1}) is also satisfied at the leading perturbative order. This is the main result of the work.

A few remarks are in order.

Firstly, although the above calculation is simple and straightforward, it would not have been possible without two conditions: firstly, the fundamental gauge condition (\ref{def.hf2}), which largely fixes the ansatz needed to achieve (\ref{boxhuv}); and secondly, the special relation (\ref{def.hf1}), without which (\ref{eq.huv1}) would have nothing to do with (\ref{eq.Tuv1}).

Secondly, although the above calculation is perturbative on the gravity side, (\ref{eq.Tuv1}) is still a nonlinear equation for hidden fluid due to the intriguing fact that the velocity of hidden fluid is not controlled by the magnitude of \acp{GW}, but can be large no matter how weak the \acp{GW} are.

Thirdly, it is possible to write the dissipative term $\cX^{\mu\nu}$ in (\ref{def.cX}) in terms of other hydrodynamical quantities of hidden fluid,
%%%
\bea\cX_{\mu\nu}&=&\kappa(\pd^\bot_\mu\cU_\nu+\pd^\bot_\nu\cU_\mu) +\lambda\Pi_{\mu\nu}(\pd\cdot\cU)\nn\\
&&+\zeta^\bot_\mu\cW^\bot_\nu+\zeta^\bot_\nu\cW^\bot_\mu\,,\label{def.cY}\eea
%%%
where $\cW^\rho\equiv\frac12 \varepsilon^{\rho\mu\nu} (\pd_\mu\cU_\nu-\pd_\nu\cU_\mu)\,$, $O^\bot_\mu\equiv \Pi_\mu^\nu O_\nu\,$ with $O_\mu\in\{\pd_\mu\,$, $\zeta_\mu\,$, $\cW_\mu\}\,$, and $\kappa\,$, $\lambda$ and $\zeta_\mu$ are free coefficients. There are only five free functions in (\ref{def.cY}) but the equation contains six constraints, so to make the equation work one must place an extra constraint on the order-1 coefficients. Remarkably, this extra constraint can be placed on $F_{20}\,$, which is unconstrained so far, and so the general \ac{GW} solution in (\ref{metric.s0}) is not affected.

Last but not least, the leading order solution (\ref{metric.s0}) only contains functions that depend on ($\mu,\theta,\phi$). So the allowed leading order variation of hidden fluid is along a codimension-one hypersurface, which is reminiscent of the hypersurfaces needed in various duality treatment mentioned at the beginning of the paper. This indicates that the present construction might have some connection to the duality treatment.

To better understand the nature of the present construction, one can consider allowing the order-1 coefficients to depend on all four coordinates, ($\mu,r,\theta,\phi$), then (\ref{GW.s1}) and (\ref{metric.s0}) no longer give a valid solution, nor is (\ref{boxhuv}) or all the discussions that follow from it. An inspection of the equations shows that the problem is due to the variation of metric elements along the $r$-direction at a scale much smaller than $r_0\,$. Such variation can be caused by local perturbations but not by the \ac{GW} sources at an $r_0$ distance away (whose effect is already included through the dependence on $u$). So in the present construction, (\ref{eq.Tuv1}) is only valid for \ac{GW} radiations in regions near the null infinity. It would be interesting to see if there is a way to make (\ref{eq.Tuv1}) work for other spacetime solutions.

\section{Conclusion}\label{sec:con}

In this work, we have studied the properties of hidden fluid underlying the general \ac{GW} spacetime near the null infinity. Although some peculiar features have been found for the velocity of hidden fluid, the properties of hidden fluid still appear to be physically possible.

More importantly, we find that hidden fluid underlying the \ac{GW} spacetimes near null infinity indeed satisfies the relativistic hydrodynamic equations (\ref{eq.Tuv1}) in the Minkowski background. Although such a result fits well with the major requirement of the fluid/gravity equivalence, it comes as a total surprise because it has been thought to be mathematically difficult to achieve.

Such result does not necessary mean that the hidden fluid model is correct, but it does provide a concrete example satisfying both of the major requirements ({\bf R1} and {\bf R2} in Section \ref{sec:intro}) expected for the fluid/gravity equivalence, and so it can serve as a good starting point for further studies on the problem.

A concrete picture about the underlying structure of spacetime, such as hidden fluid, can help us better understand how \ac{GR} might be modified. For example, one may now take a fresh look at some of the modified gravity theories involving an additional vector field, with the Einstein-aether gravity \cite{Jacobson:2007veq} being a possible good example \cite{Mei:2022ksw}. Such effort should be meaningful given the many new opportunities for testing \ac{GR} offered by the breakthroughs in \ac{GW} detection \cite{LIGOScientific:2016aoc,LIGOScientific:2016lio,LIGOScientific:2019fpa,LIGOScientific:2020tif} and the prospect of many future \ac{GW} detectors \cite{Punturo:2010zz,Maggiore:2019uih,Reitze:2019iox,LISA:2017pwj,TianQin:2015yph,Gong:2021gvw,Kawamura:2020pcg,Berti:2015itd}.

{
\begin{acknowledgments}
This work has been supported in part by the Guangdong Major Project of Basic and Applied Basic Research (Grant No. 2019B030302001).
\end{acknowledgments}
}

%%%%%%%%%%%%%%%%%%%%%%%%%%%%%%%%%%%%%%%%%%%%%%%%%%%%%%%%%%%%%%%%
%\bibliographystyle{unsrt}
\bibliography{hfgw}
%%%%%%%%%%%%%%%%%%%%%%%%%%%%%%%%%%%%%%%%%%%%%%%%%%%%%%%%%%%%%%%%
\end{document}